\documentclass[letterpaper]{article} 
\usepackage{aaai2026}  
\usepackage{times}  
\usepackage{helvet}  
\usepackage{courier}  
\usepackage[hyphens]{url}  
\usepackage{graphicx} 
\usepackage{natbib}  
\usepackage{caption} 
\usepackage{algorithm}
\usepackage{enumitem}
\usepackage{algorithmic}
\usepackage{amsmath}
\usepackage{makecell}

\usepackage{newfloat}
\usepackage{listings}
\DeclareCaptionStyle{ruled}{labelfont=normalfont,labelsep=colon,strut=off} 
\floatstyle{ruled}
\newfloat{listing}{tb}{lst}{}
\floatname{listing}{Listing}
\title{Long document summarization using page specific target text alignment and distilling page importance}
\author {
    Pushpa Devi
    , Ayush Agrawal
    , Ashutosh Dubey
    , C. Ravindranath Chowdary
}
\affiliations {
    Department of Computer Science and Engineering,\\Indian Institute of Technology
(BHU) Varanasi-221005, India\\
    \{pushpadevi.rs.cse23, ayush.agrawal.cse20, ashutosh.dubey.cse20, rchowdary.cse\}@iitbhu.ac.in
}

\usepackage{bibentry}

\begin{document}

\maketitle

\begin{abstract}
The rapid growth of textual data across news, legal, medical, and scientific domains is becoming a challenge for efficiently accessing and understanding large volumes of content. It is increasingly complex for users to consume and extract meaningful information efficiently. Thus, raising the need for summarization. Unlike short document summarization, long document abstractive summarization is resource-intensive, and very little literature is present in this direction. BART is a widely used efficient sequence-to-sequence (seq-to-seq) model. However, when it comes to summarizing long documents, the length of the context window limits its capabilities. We proposed a model called \textbf{PTS} (Page-specific Target-text alignment Summarization) that extends the seq-to-seq method for abstractive summarization by dividing the source document into several pages. PTS aligns each page with the relevant part of the target summary for better supervision. Partial summaries are generated for each page of the document. We proposed another model called PTSPI (Page-specific Target-text alignment Summarization with Page Importance), an extension to PTS where an additional layer is placed before merging the partial summaries into the final summary. This layer provides dynamic page weightage and explicit supervision to focus on the most informative pages. We performed experiments on the benchmark dataset and found that PTSPI outperformed the SOTA by 6.32\% in ROUGE-1 and 8.08\% in ROUGE-2 scores.
\end{abstract}

\begin{links}
    \link{Datasets}{https://github.com/armancohan/long-summarization}
\end{links}

\section{Introduction}\label{sec:introduction}

Abstractive text summarization automatically generates a brief and accurate version of a text document, retaining its main ideas and general meaning. Summarization reduces the reading burden while preserving the core content and, at the same time, enabling faster comprehension and efficient decision-making. Most summarization systems are extractive, which selects and combines a few important sentences or phrases from the original text. On the contrary, abstractive summarization systems generate new sentences that convey the core ideas more naturally and human-like.
 
Memory size is the prime barrier for summarizing a long document because self-attention models require quadratic memory growth with respect to the source document. Automatic summarization of long documents, such as scientific articles, technical reports, and legal briefs, remains challenging in natural language processing (NLP). Although transformer-based models such as BART~\cite{lewis-etal-2020-bart} and PEGASUS~\cite{pegasus-zhang-2020}have achieved remarkable results on short texts, their fixed input length and quadratic attention complexity preclude direct application to texts exceeding a few thousand tokens~\cite{lewis-etal-2020-bart,pegasus-zhang-2020}. Efficient attention mechanisms ( Longformer~\cite{beltagy-LED-LONGFORMER}, BigBird~\cite{Zaheer}) and hierarchical architectures (e.g., HAT-BART~\cite{HATBART-rohde2021hierarchical}, PRIMERA~\cite{xiao-etal-2022-primera}) partially mitigate these issues, but often sacrifice global coherence or fine-grained locality.

PageSum~\cite{Pagesum} addresses this gap by segmenting a document into fixed-length ``pages" and processing each page independently with a pretrained encoder–decoder. It aggregates page-level representations via a learned confidence layer. This locality-aware method reduces memory requirements while preserving intra-page expressivity. However, PageSum relies solely on the downstream summarization loss to infer page importance. It assigns the entire reference summary to every page during training and lacks explicit document structure modeling. PageSum and its predecessors generated a summary for each page by feeding the entire target summary to every page. This hinders the quality of the produced summary. PTS and PTSPI address this problem. PTSPI applies the teacher-student architecture for distilling page importance. The model proposed in ~\cite{Pagesum} also utilized locality in summarizing the text document, which works on the principle of ``locality", one of the main principles of virtual memory systems~\cite{denning2005locality}. Authors in ~\cite{long-doc-efficient, Zaheer} utilized the efficient attention-based methods~\cite{Zaheer,reformer,tay2022efficient} to reduce the quadratic memory requirements for the long document summarization. Our goal is to reduce the memory requirement for long documents and give importance to the pages containing more sentences that are highly similar to the target summary.

Our main contributions are:
\begin{itemize}
\item A novel approach that aligns the relevant part of the target summary to each page instead of aligning the entire summary for each page.
\item A teacher-student architecture for distilling page importance, which provides explicit supervision for the model to focus more on the most informative pages.
\end{itemize}
These enhancements yield a robust, interpretable, and scalable approach to long-document abstractive summarization. On the arXiv~\cite{Dataset-ARXIV} benchmark dataset, our both model surpasses the state-of-the-art ROUGE scores, demonstrating substantial gains in informativeness and coherence over both PageSum and other baselines.

The remainder of this paper is organized as follows. Section~\ref{sec:related} reviews related work on the summarization task. In Section~\ref{sec:problem-formulation}, we discussed problem formulation. Section~\ref{sec:method} details our integrated model architecture, including page-specific target text alignment (PTS) and distillation approach (PTSPI). Section~\ref{sec:experimental Setup} describes the datasets, SOTA, and experimental details. Finally, Section~\ref{sec:conclusion} concludes and outlines future directions.

\section{Related work}\label{sec:related}
Natural language processing tasks such as Automatic summarization systems (ATS) are becoming more prominent in domains such as NEWS, scientific, medical, etc. ATS aims to provide a succinct summary of the document. Based on the nature of the output generated by ATS, they are categorized as extractive ~\cite{WitbrockM99,knight2000statistics, radev1998generating} and abstractive summarization ~\cite{radev2004centroid, conroy2005classy}

The extractive summarization model selects the salient snippets from the input document and rearranges them to produce the summary. Extractive summarization is categorized as unsupervised and supervised methods. The popular unsupervised extractive methods, which are graph-based, textRank~\cite{mihalcea2004textrank} and Lexrank~\cite{erkan2004lexrank}, are very prominent. Authors in ~\cite{ZhengL19,dong-2021-discourse,liang2021improving} developed unsupervised extractive long document summarization models. Few earlier works were based on RNN \cite{Extract-super1, Nallapati-2017}  and BERT-based models ~\cite{extra-super2,extract-super-bert-dmn,extract-super3-graph}, which were supervised extractive models, and a few are based on large language models ~\cite {leveraginglarge}, GoSum~\cite{gosum} is an extractive approach that applies a graph-based model and reinforcement learning techniques for summarizing long documents. They construct a heterogeneous graph that represents each document at various discourse levels.

An abstractive summarization model understands the contextual information of the source document and creates a summary just as humans do. Various abstractive methods are rule-based and neural Network-based. In rule-based methods, handpicked rules and categories are given to the system to find the meaningful candidates, which are then used to create the final summary. In contrast, Neural Network-based models, the semantics of the inputs are encoded by the encoder, and the decoder module decodes to generate the summary; these are known as encoder-decoder models~\cite{Seq-to-seq}.

The early neural-network-based approaches utilized the RNN and LSTM cells~\cite{lstm} in their neural network to perform encoding-decoding of the input documents. Bahdanau et al. (2015) developed the attention mechanism to help the decoder focus only on the relevant part of the input. Based on the Bahdanau attention mechanism, many summarization models were developed. One of the popular summarization models was developed by~\cite{rush-2015-neural-attention}; they were the first to utilize the attention mechanism for summarization. ~\cite{see2017get} combine abstractive generation with a copying mechanism, allowing the model to generate new words and copy from the text. Despite these advancements, abstractive non-transformer-based models were still limited in coherence and fluency. Recent transformer-based architectures address these shortcomings, but these models are more suitable for processing short documents only.

\subsection{Transformer-based approaches}

The initial transformer-based models use an extract-then-abstract strategy; they viewed Wikipedia article generation as a multi-document summarization task. The models by ~\cite{BertSum-liu-lapata-2019-text,lewis-etal-2020-bart,pegasus-zhang-2020}  for summarizing the documents are based on the transformer model ~\cite{vaswani2017attention}. All these models work best for summarizing short documents.

To address the problem of handling long sequences, a few methods are available, including efficient attention-based models (~\cite{beltagy-LED-LONGFORMER, Zaheer} and a few hierarchical architecture-based models developed by  ~\cite{HATBART-rohde2021hierarchical,xiao-etal-2022-primera}. The self-attention mechanism of the transformer models hinders their applicability over long documents, as they require quadratic memory as the input grows. Also, most transformer models often sacrifice global coherence or fine-grained locality. To address this issue, authors in  ~\cite{beltagy-LED-LONGFORMER,manakul-gales-2021-loBART, poolingformer} adopted fixed pattern attention on some local context. Reformer ~\cite{reformer} and Sinkhorn attention~\cite{sink-horn-attention} try to learn the attention patterns. HEPOS~\cite{Hepos-huang-efficient} modifies the encoder-decoder attention using head-wise positional strides to pinpoint the salient information from the source document.

DANCER developed by~\cite{dancer} proposes a divide and conquer based abstractive summarization approach. It divides a long document and summaries into segments based on the ROUGE scores, and combines the output from different sections to form the final summary. Authors in~\cite{text-align-UNBALANCED-OPTIMAL-TRANSPORT} propose a model that jointly learns the alignment of the number of summary sentences for each section using unbalanced optimal transport theory. An empirical survey on long document summarization by ~\cite{Koh_2022-empirical-survey} discovers that most of the sentences in the references are assigned to the introduction and conclusion sections, and only 35\% aligned with results and methods. Other methods based on hybrid and LLM based ~\cite{Lodoss,mao-etal-2022-dyle, rstlora-24, hera-25, Merella-lapata-2025-context-llm} are also devised for summarizing long documents.

\section{Problem Formulation}\label{sec:problem-formulation}
Let $D = \{w_1, w_2, ..., w_N\}$ denote a long document consisting of $N$ tokens. The long-document text summarization aims to generate a concise summary $S = \{y_1, y_2, ..., y_M\}$, with $M \ll N$, such that summary $S$ captures the salient features of $D$.

The quadratic memory requirement of standard transformer models on long input documents hinders their utilization. So a segmentation-based approach is proposed in literature~\cite{Dataset-ARXIV} that is to partition the document $D$ into $P$ non-overlapping segments referred to as \emph{pages}, each containing up to $L$ tokens.
\begin{equation}
    D = \{P_1, P_2, ..., P_P\}, \quad P_j = \{w_{a_j}, ..., w_{b_j}\}, \quad 1 \leq j \leq P
\end{equation}
Where $a_j$ and $b_j$ are the start and end indices of tokens in page $P_j$.
In Liu et al. 2022, each page $P_j$ is independently encoded and decoded to produce a page-level representation $h_j$. These hidden representations of page summaries are aggregated to form the basis for final summary generation. However, this partitioning introduces few fundamental challenges:
 \begin{enumerate}[label=(\roman*)]
 \item  \textbf{imprecise Target summary to each page:} We hypothesize that independent encoding and decoding can not produce a good-quality summary without a relevant target text alignment. This is evident from the performance of the PageSum model~\cite{Pagesum}, which provides the entire target summary to each page.
 \item \textbf{Imprecise Page Importance:} Generally, few of the pages contain more salient information than others. Summarization loss alone may not be sufficient for the model to infer the relevancy of a page to the summary. Hence, we hypothesize that explicit supervision is desirable.
 \end{enumerate}

Let $f_\theta$ denote the neural summarization model parameterized by $\theta$. The model takes as input the set of page representations $\{h_1, ..., h_P\}$ and generates the summary $S$:
\begin{equation}
S = f_\theta(h_1, ..., h_P)
\end{equation}

In the PageSum model~\cite{Pagesum}, the aggregation by confidence layer produces static weights $\{c_1, ..., c_P\}$ for the pages. The final summary token at each step is generated from a weighted sum of the page representations:
\begin{equation}\label{}
  p(y_i \mid y_{l<i}, D) = \mathrm{softmax}\left( W \left( \sum_{j=1}^P c_j \cdot h_j \right) + b \right)  
\end{equation}

where $W$ and $b$ are learnable parameters.

\textbf{Problem Statement:} \emph{Given a long document $D$ partitioned into pages $\{P_1, ..., P_P\}$, design a summarization model that (i) aligns the page to relevant target text and (ii) dynamically learns the importance of each page.}

\section{Proposed Method}\label{sec:method}
Our novel contribution enables the model to generate informative and concise summaries. Also, the summaries are structurally coherent and explicitly grounded in the most relevant parts of the document.

\subsection{Page-Specific Relevant Target-Text Alignment (PTS)}\label{page-specific-target-alignment}
\begin{figure*}[htbp!]
  \centering
  \includegraphics[width=\textwidth]{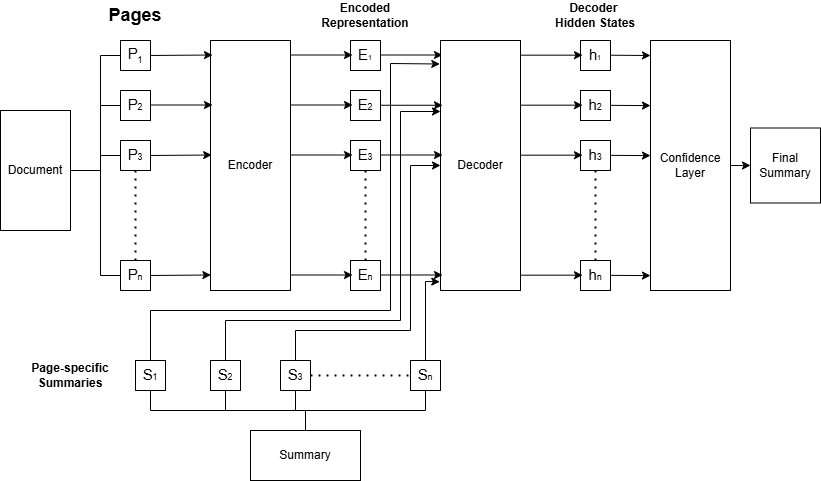}
   \caption{Architecture of PTS model. Decoder will get Page-specific target summary sentences only.} 
  \label{fig:architecture_pageSumRef}
\end{figure*}

We designed and developed a novel hierarchical summarization model \textbf{PTS}, suitable for summarizing long documents. PTS avoids quadratic growth of memory by considering each document as a series of non-overlapping pages with fixed-page length, and each page will get the relevant part of the gold summary.
Instead of supplying the entire reference summary to every page, we align each summary sentence \(s\) to its most relevant page. Figure~\ref{fig:architecture_pageSumRef} depicts the overall architecture for summarizing the documents using page-specific summary sentences only.

In the original PageSum, the complete reference summary is provided to every page during training, regardless of the actual relevance of each summary sentence to a given page. A complete reference summary to each page can introduce noise and reduce learning precision. We address the imprecise alignment of the summary sentence by assigning it to the most semantically relevant page using cosine similarity between the summary sentence and page embeddings.

\text{Let } $e(s_k)$, $e(P_j)$ $\in {R}^k$ be the embedding (BART) of the summary sentence and page, respectively. For each summary sentence $s_k$, we compute its cosine similarity to every page $P_j$ and assign $s_k$ to the page with the highest similarity,
\begin{equation}
P_j = \mathop{\mathrm{argmax}}\limits_{j} \, \text{sim}\left( e(P_j), e(s_k) \right)
\end{equation}
\begin{equation}
    \text{sim}(\mathbf{e}(P_j), \mathbf{e}(s_k) ) = \frac{\mathbf{e}(P_j) \cdot\mathbf{e}(s_k) }{ \|\mathbf{e}(P_j)\|\|\mathbf{e}(s_k)\|}\label{cosine_similarity}
\end{equation}
Eq.~\ref{cosine_similarity} is cosine similarity between the embedding of the page and summary sentences.

Aligning pages with relevant target sentences yields page-specific reference summaries $\{S_j\}$, where $S_j$ contains only those summary sentences most relevant to page $P_j$. During training, decoding each page using only its assigned summary sentences results in more precise and context-sensitive learning. During training, the decoder for \(P_j\) generates \(S_j\) rather than the full abstract.

\subsection{Dynamic page importance using Teacher–Student distillation}\label{teacher-student-distillation}

\begin{figure}[htbp!]
  \centering
  \resizebox{1\columnwidth}{!}{
  \includegraphics[width=\linewidth]{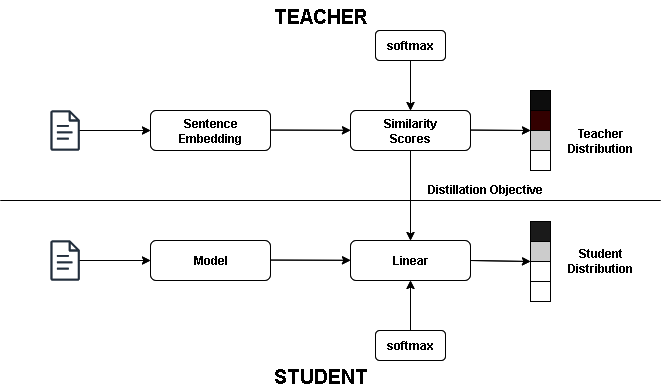}}
  \caption{Teacher–student distillation mechanism for learning page importance weights.}
  \label{fig:teacher_student}
\end{figure}

In Figure~\ref{fig:teacher_student}, our dynamic page importance approach provides a supervisory signal for page salience via a teacher–student paradigm~\cite{distillingknowledgeneuralnetwork,liu2021noisy-distill}:
\begin{itemize}
  \item \textbf{Teacher distribution Generation:} Each document \(D\)\ containing \(P_1,\dots,P_P\) pages, we generate a provisional summary for each page \(P_j\). We embed both the page summary and the gold summary using a frozen sentence encoder (e.g., Sentence-BERT\cite{Reimers2019SBERT}) and compute cosine similarities \(\{\alpha_j\}\)\ . We normalize the similarities using softmax to obtain the teacher distribution. 
  \begin{equation}
      T_j = \frac{\exp(\alpha_j)}{\sum_{k}\exp(\alpha_k)} 
  \end{equation}
 
  \item \textbf{Student Prediction: } At each decoding step, the model dynamically generates a projected distribution \(Z_j\) over pages based on the number of sentences, and each page's weightage is calculated. Pages with a larger number of sentences get high weightage, and those with fewer sentences get less weightage.
  
  \item \textbf{Distillation Loss:} We align student distribution with teacher distribution via KL-divergence:
  \begin{equation}
       \mathcal{L}_{\mathrm{KL}} = \sum_{j=1}^n T_j \log\frac{T_j}{Z_j}\
  \end{equation}
\end{itemize}
\subsection{Page-specific Target-text alignment Summarization with Page Importance (PTSPI)}
 PTSPI is an integration of models proposed in Section~\ref{page-specific-target-alignment} and~\ref{teacher-student-distillation}, PTSPI utilizes page-specific relevant target text alignment cross-entropy loss and importance weights calculated by KL loss to generate summaries for each page. Final loss as:
\begin{equation}
    \mathcal{L} = (1-\lambda)\,\mathcal{L}_{\mathrm{XENT}} + \lambda\,\mathcal{L}_{\mathrm{KL}}\,
\end{equation}
where  \(\lambda\) ranges between 0 to 1. We took \(\lambda\)=0.1 weight the distillation loss.
At inference (last layer in Figure~\ref{fig:architecture_pageSumRef}), the confidence layer’s predictions \(ZZ_j\) determine the weightage of each page’s contribution. These confidence layer predictions are used to combine the local hidden states of each page to produce the final aggregated summary.

\section{Experimental setup}\label{sec:experimental Setup}
\subsection{Datasets and Metrics}\label{subsec:Dataset}
For empirical evaluation, we used the widely adopted \textbf{arXiv Summarization Dataset}\footnote{https://github.com/armancohan/long-summarization} developed by ~\cite{Dataset-ARXIV}, which is designed explicitly for benchmarking long-document summarization models. This dataset comprises scientific research articles from the arXiv preprint server, each paired with its abstract. This corpus is particularly challenging due to the documents’ substantial length and complex structure, with an average article length of 8613 tokens and summaries (abstracts) averaging around 362 tokens. 

The dataset is partitioned into standard Train/Validation/Test sets of 203,037/6,436/6,440 documents, respectively. Each instance contains two fields: article (the full text of the paper) and abstract (the reference summary). Before training, all documents are preprocessed by removing LaTeX commands, normalizing whitespace, and segmenting each article into non-overlapping pages of up to 1024 tokens to comply with the input constraints of the
underlying transformer models. This segmentation enables efficient batch processing.

\subsection{State-of-the-art models}\label{Baseline}
To check the effectiveness of our novel model, we compare it with several baselines devised for summarizing long documents:
\begin{enumerate}

    \item \textbf{LED (Longformer Encoder-Decoder)} introduce sparse attention. This attention is combined with encoder-decoder attention to process the long documents for maintaining the long-context sensitivity. LED~\cite{beltagy-LED-LONGFORMER} extends the standard encoder-decoder architecture by incorporating sparse self-attention in the encoder.

    \item \textbf{HEPOS}~\cite{Hepos-huang-efficient} employs a hybrid attention mechanism that integrates efficient self-attention with encoder-decoder attention. This model optimizes long-document summarization, with a noticeable balance between computational efficiency and representational power.

    \item \textbf{PRIMERA} by \cite{xiao-etal-2022-primera}  builds upon the LED architecture and enhances it through task-specific pretraining for multi-document summarization. This additional pretraining enables the model to better capture relationships across different segments of lengthy documents.

    \item \textbf{HAT-BART}~\cite{HATBART-rohde2021hierarchical}  is based on the original BART model \cite{lewis-etal-2020-bart}. HAT-BART introduces hierarchical attention layers to model sentence-level interactions. Unlike LED and HEPOS, HAT-BART employs full attention throughout the architecture, which can improve performance at the cost of higher computational overhead.

    \item \textbf{PageSum} Proposed model that leverages locality in abstractive summarization for long document~\cite{Pagesum}. Their model utilizes independent encoding and decoding of each page, but the entire reference summary was given to each page at decoding.
\end{enumerate}

\subsection{Implementation details}\label{subsec:implementaion}
We implemented PTS and PTSPI using the HuggingFace Transformers library\footnote{ https://github.com/huggingface/transformers}, with BART as the encoder-decoder backbone. We used BART checkpoints pretrained on the CNN/DailyMail summarization dataset for all approaches, and segmented the document into non-overlapping pages of up to 1024 tokens. Page-specific reference summaries are assigned during training using cosine similarity between page and summary sentence embeddings. Models are trained for three epochs on 200,000 training samples, with early stopping based on validation \textbf{Bert-score}. We keep the learning rate at 2e-3 and the batch size at three samples.

The total loss comprised the standard cross-entropy loss (XENT) for summary generation and an auxiliary KL divergence loss for teacher–student page importance distillation, weighted by a tunable hyperparameter $\lambda$=0.1.

We assessed model performance using ROUGE-1 and ROUGE-2 F1 scores~\cite{lin-2004-rouge}, computed on the test set. These metrics quantify the overlap between generated summaries and human-written abstracts at the unigram, bigram, and longest common subsequence levels. All experiments are conducted on high-performance computational resources to accommodate long-document summarization’s substantial memory and processing requirements. We utilize the PARAM Shivay\footnote{https://www.iitbhu.ac.in/cf/scc} supercomputer at IIT (BHU), Varanasi, a part of the National Supercomputing Mission. Specifically, we leveraged two GPU nodes, each with two NVIDIA cards. Each of these cards is 16 GB. We implemented our model by distributing BART-large’s encoder and decoder modules across the two nodes, enabling efficient training and inference on large-scale data. Each epoch required approximately 5.5 days to complete on the two-node PARAM Shivay setup. The additional overhead from dynamic page-summary alignment and teacher–student distillation was accomplished through parallel data loading and efficient GPU utilization. The final model checkpoint was selected based on the best validation \textbf{Bert-score}.

\section{Results and Analysis }\label{results}

Table~\ref{table 1} shows the performance on the arXiv dataset. When we align all pages to their relevant reference summaries, PTS outperforms the state-of-the-art in ROUGE-1 score by 3.42\% and ROUGE-2 by 5.60\%. Similarly, along with relevant target text-alignment when the weightage to each of the partial summaries is learnt dynamically before computing the final summary, PTSPI outperforms the state-of-the-art in ROUGE-1 score by 6.32\% and ROUGE-2 by 8.08\%. In \cite{fabbri-etal-2021-summeval}, the authors observe that higher ROUGE-2 scores correlate with human-rated coherence. Our hypothesis in Section~\ref{sec:problem-formulation} is true, as evident by the data in Table~\ref{table 1}.

We analyze two important factors:
\begin{enumerate}
   \item Reference of target text alignment: To identify a suitable metric for aligning the relevant summary sentences to each document page, we initially experimented on 20,000 random samples for a single epoch. We compared R1, R2, R-L and BERTscore metrics and empirically found that BERTscore suits this task. The Table~\ref{t2} shows the performance of various metrics.
    \item Complexity: Overall space complexity of PTS is linear. We have divided the document into $\text{at most } \left\lfloor \frac{{D_l}}{P_l} \right\rfloor$ pages. The space complexity of the encoder self-attention for one page is $O\left(P_l^2\right)$. Therefore, the complexity of all the pages is $O\left({P_l^2} \cdot \frac{D_l}{P_l}\right)$\ . PTS incurs additional overhead of aligning sentences to each page. So we needed higher computational power. To overcome this problem, we use model parallelism and split the model across two GPU nodes, each with 32GB of memory. Training time per epoch for each model is shown in Table~\ref{table 3}. Part of the experiments (pagesum*) were conducted on AWS g6dn.12xlarge instances equipped with 4× NVIDIA L4 Tensor Core GPUs with 24 GB of memory.
\end{enumerate}
\begin{table}[htbp!]
    \centering
     \resizebox{.95\columnwidth}{!}{
    \begin{tabular}{lcc}
        \hline
        \textbf{Model} & \textbf{ROUGE-1} & \textbf{ROUGE-2} \\
        \hline
        LED~\cite{beltagy-LED-LONGFORMER} & 48.10 & 19.78 \\
        HEPOS~\cite{Hepos-huang-efficient} & 48.24 & 20.26 \\
        PRIMERA~\cite{xiao-etal-2022-primera} & 47.60 & 20.80 \\
        HAT-BART~\cite{HATBART-rohde2021hierarchical} & 46.68 & 19.07 \\
        PageSum~\cite{Pagesum} & 49.72 & 21.06 \\
        PageSum* & 48.54 & 20.55 \\
        \textbf{PTS} & \textbf{50.20} & \textbf{21.70} \\
        \textbf{PTSPI} & \textbf{51.61} & \textbf{22.21} \\
    \hline
    \end{tabular}  }
    \caption{Comparison of state-of-the-art models for long-document summarization on the arXiv dataset. PageSum* refers to our reimplementation of the PageSum model trained on 200,000 samples for 3 epochs.}
    \label{table 1}
\end{table}
 \begin{table}[htbp!]
    \centering
    \resizebox{.9\columnwidth}{!}{
    \begin{tabular}{c c c}
        \hline
        \textbf{Similarity Metric} & \textbf{ROUGE-1} & \textbf{ROUGE-2} \\
        \hline
        ROUGE-1 & 44.10 & 18.64 \\
        ROUGE-2 & 44.63 & 18.76 \\
        ROUGE-L & 45.09 & 18.92 \\
        \textbf{BERTScore} & \textbf{45.43} & \textbf{19.24} \\
        \hline
    \end{tabular}
    }
    \caption{Comparison of sentence-to-page alignment metrics based on ROUGE performance. Evaluation was conducted on 20,000 samples for a single epoch.}
    \label{t2}
\end{table}
\begin{table}
     \centering
        \resizebox{.99\columnwidth}{!}{
        \begin{tabular}{lll}
        \hline
        \textbf{Model} & \textbf{Training time} & \textbf{System configuration}\\
        \hline
        PageSum* & 60 hours & 4× NVIDIA L4 Tensor Core GPUs with 24 GB of memory\\
        PTS & 108 hours & 2 GPUs each with 2 NVIDIA cards of 16GB memory\\
        PTSPI & 132 hours & 2 GPUs each with 2 NVIDIA cards of 16GB memory\\
    \hline
    \end{tabular}  
    }
    \caption{Training time taken by models per epoch}
    \label{table 3}
\end{table}
\section{Conclusion}\label{sec:conclusion}
We proposed PTS and PTSPI models for summarizing long documents. PTS assigns reference summary sentences to the semantically relevant page segments. Utilizing BERTScore, a similarity metric grounded in contextual embeddings, we proposed a robust alignment strategy that dynamically maps summary sentences to their contextually similar pages. Also, PTSPI adds weightage to each page-wise summary. While generating the final summary, the pages will contribute proportionally to their weightage. This page's importance is computed using the loss between the teacher-student distribution of pages. The experimental results show that PTSPI outperforms the SOTA by a significant margin. We compare PTS and PTSPI with the existing efficient attention and locality-based models, and experimentally, we show that both models are superior. In future, we would like to further enhance the model by considering many-to-many alignments and cross-page dependency modeling, which can significantly elevate the quality, coherence, and informativeness of long-document summarization.
\section{Acknowledgments}
We sincerely thank the IIT(BHU) Supercomputing Center facilities for providing the high-performance computing resources required for this work. We also acknowledge and thank the support of AWS Aakash Ganga IIT (BHU), for the cloud infrastructure that significantly helped this research's progress. Their support was invaluable in enabling the efficient execution of this research work.

\bibliography{references}

\begin{thebibliography}{52}
\providecommand{\natexlab}[1]{#1}

\bibitem[{Beltagy, Peters, and Cohan(2020)}]{beltagy-LED-LONGFORMER}
Beltagy, I.; Peters, M.~E.; and Cohan, A. 2020.
\newblock Longformer: The Long-Document Transformer.
\newblock \emph{CoRR}, abs/2004.05150.

\bibitem[{Bian et~al.(2024)Bian, Huang, Zhou, Huang, and Zhu}]{gosum}
Bian, J.; Huang, X.; Zhou, H.; Huang, T.; and Zhu, S. 2024.
\newblock GoSum: extractive summarization of long documents by reinforcement learning and graph-organized discourse state.
\newblock \emph{Knowl. Inf. Syst.}, 66(12): 7557--7580.

\bibitem[{Cho et~al.(2022)Cho, Song, Wang, Liu, and Yu}]{Lodoss}
Cho, S.; Song, K.; Wang, X.; Liu, F.; and Yu, D. 2022.
\newblock Toward Unifying Text Segmentation and Long Document Summarization.
\newblock In Goldberg, Y.; Kozareva, Z.; and Zhang, Y., eds., \emph{Proceedings of the 2022 Conference on Empirical Methods in Natural Language Processing}, 106--118. Abu Dhabi, United Arab Emirates: Association for Computational Linguistics.

\bibitem[{Cohan et~al.(2018)Cohan, Dernoncourt, Kim, Bui, Kim, Chang, and Goharian}]{Dataset-ARXIV}
Cohan, A.; Dernoncourt, F.; Kim, D.~S.; Bui, T.; Kim, S.; Chang, W.; and Goharian, N. 2018.
\newblock A Discourse-Aware Attention Model for Abstractive Summarization of Long Documents.
\newblock In \emph{Proceedings of the 2018 Conference of the North {A}merican Chapter of the Association for Computational Linguistics: Human Language Technologies, Volume 2 (Short Papers)}. New Orleans, Louisiana: Association for Computational Linguistics.

\bibitem[{Conroy, Schlesinger, and Stewart(2005)}]{conroy2005classy}
Conroy, J.~M.; Schlesinger, J.~D.; and Stewart, J.~G. 2005.
\newblock CLASSY query-based multi-document summarization.
\newblock In \emph{Proceedings of the 2005 Document Understanding Workshop, Boston}. Citeseer.

\bibitem[{Cui and Hu(2021)}]{extract-super-bert-dmn}
Cui, P.; and Hu, L. 2021.
\newblock Sliding Selector Network with Dynamic Memory for Extractive Summarization of Long Documents.
\newblock In \emph{Proceedings of the 2021 Conference of the North American Chapter of the Association for Computational Linguistics: Human Language Technologies}, 5881--5891. Online: Association for Computational Linguistics.

\bibitem[{Cui, Hu, and Liu(2020)}]{extract-super3-graph}
Cui, P.; Hu, L.; and Liu, Y. 2020.
\newblock Enhancing Extractive Text Summarization with Topic-Aware Graph Neural Networks.
\newblock In \emph{Proceedings of the 28th International Conference on Computational Linguistics}, 5360--5371. Barcelona, Spain (Online): International Committee on Computational Linguistics.

\bibitem[{Denning(2005)}]{denning2005locality}
Denning, P.~J. 2005.
\newblock The locality principle.
\newblock \emph{Communications of ACM}, 48(7): 19--24.

\bibitem[{Dong, Mircea, and Cheung(2021)}]{dong-2021-discourse}
Dong, Y.; Mircea, A.; and Cheung, J. C.~K. 2021.
\newblock Discourse-Aware Unsupervised Summarization for Long Scientific Documents.
\newblock In \emph{Proceedings of the 16th Conference of the European Chapter of the Association for Computational Linguistics: Main Volume}, 1089--1102. Online: Association for Computational Linguistics.

\bibitem[{Erkan and Radev(2004)}]{erkan2004lexrank}
Erkan, G.; and Radev, D.~R. 2004.
\newblock Lexrank: Graph-based lexical centrality as salience in text summarization.
\newblock \emph{Journal of artificial intelligence research}, 22: 457--479.

\bibitem[{Fabbri et~al.(2021)Fabbri, Kry{\'s}ci{\'n}ski, McCann, Xiong, Socher, and Radev}]{fabbri-etal-2021-summeval}
Fabbri, A.~R.; Kry{\'s}ci{\'n}ski, W.; McCann, B.; Xiong, C.; Socher, R.; and Radev, D. 2021.
\newblock {S}umm{E}val: Re-evaluating Summarization Evaluation.
\newblock \emph{Transactions of the Association for Computational Linguistics}, 9: 391--409.

\bibitem[{Gidiotis and Tsoumakas(2020)}]{dancer}
Gidiotis, A.; and Tsoumakas, G. 2020.
\newblock A Divide-and-Conquer Approach to the Summarization of Long Documents.
\newblock \emph{IEEE/ACM Transactions on Audio, Speech, and Language Processing}, 28: 3029--3040.

\bibitem[{Hemamou and Debiane(2024)}]{leveraginglarge}
Hemamou, L.; and Debiane, M. 2024.
\newblock Scaling Up Summarization: Leveraging Large Language Models for Long Text Extractive Summarization.
\newblock arXiv:2408.15801.

\bibitem[{Hinton, Vinyals, and Dean(2015)}]{distillingknowledgeneuralnetwork}
Hinton, G.; Vinyals, O.; and Dean, J. 2015.
\newblock Distilling the Knowledge in a Neural Network.
\newblock arXiv:1503.02531.

\bibitem[{Hochreiter and Schmidhuber(1997)}]{lstm}
Hochreiter, S.; and Schmidhuber, J. 1997.
\newblock Long Short-Term Memory.
\newblock \emph{Neural Comput.}, 9(8): 1735--1780.

\bibitem[{Huang et~al.(2021)Huang, Cao, Parulian, Ji, and Wang}]{Hepos-huang-efficient}
Huang, L.; Cao, S.; Parulian, N.; Ji, H.; and Wang, L. 2021.
\newblock Efficient Attentions for Long Document Summarization.
\newblock In \emph{Proceedings of NAACL-HLT 2021}, 1419--1436. Association for Computational Linguistics.

\bibitem[{Kitaev, Kaiser, and Levskaya(2020)}]{reformer}
Kitaev, N.; Kaiser, L.; and Levskaya, A. 2020.
\newblock Reformer: The Efficient Transformer.
\newblock In \emph{8th International Conference on Learning Representations, {ICLR} 2020, Addis Ababa, Ethiopia, April 26-30, 2020}. OpenReview.net.

\bibitem[{Knight and Marcu(2000)}]{knight2000statistics}
Knight, K.; and Marcu, D. 2000.
\newblock Statistics-based summarization-step one: Sentence compression.
\newblock \emph{AAAI/IAAI}, 2000: 703--710.

\bibitem[{Koh et~al.(2022)Koh, Ju, Liu, and Pan}]{Koh_2022-empirical-survey}
Koh, H.~Y.; Ju, J.; Liu, M.; and Pan, S. 2022.
\newblock An Empirical Survey on Long Document Summarization: Datasets, Models, and Metrics.
\newblock \emph{ACM Computing Surveys}, 55(8): 1–35.

\bibitem[{Lewis et~al.(2020)Lewis, Liu, Goyal, Ghazvininejad, Mohamed, Levy, Stoyanov, and Zettlemoyer}]{lewis-etal-2020-bart}
Lewis, M.; Liu, Y.; Goyal, N.; Ghazvininejad, M.; Mohamed, A.; Levy, O.; Stoyanov, V.; and Zettlemoyer, L. 2020.
\newblock {BART}: Denoising Sequence-to-Sequence Pre-training for Natural Language Generation, Translation, and Comprehension.
\newblock In \emph{Proceedings of the 58th Annual Meeting of the Association for Computational Linguistics}, 7871--7880. Online: Association for Computational Linguistics.

\bibitem[{Li et~al.(2025)Li, Chen, Yu, and Zhang}]{hera-25}
Li, T.; Chen, H.; Yu, F.; and Zhang, Y. 2025.
\newblock HERA: Improving Long Document Summarization using Large Language Models with Context Packaging and Reordering.
\newblock arXiv:2502.00448.

\bibitem[{Liang et~al.(2021)Liang, Wu, Li, and Li}]{liang2021improving}
Liang, X.; Wu, S.; Li, M.; and Li, Z. 2021.
\newblock Improving unsupervised extractive summarization with facet-aware modeling.
\newblock In \emph{Findings of the Association for Computational Linguistics: ACL-IJCNLP 2021}, 1685--1697.

\bibitem[{Lin(2004)}]{lin-2004-rouge}
Lin, C.-Y. 2004.
\newblock {ROUGE}: A Package for Automatic Evaluation of Summaries.
\newblock In \emph{Text Summarization Branches Out}, 74--81. Barcelona, Spain: Association for Computational Linguistics.

\bibitem[{Liu and Demberg(2024)}]{rstlora-24}
Liu, D.; and Demberg, V. 2024.
\newblock {RST}-{L}o{RA}: A Discourse-Aware Low-Rank Adaptation for Long Document Abstractive Summarization.
\newblock In \emph{Proceedings of the 2024 Conference of the North American Chapter of the Association for Computational Linguistics: Human Language, Technologies (Volume 1: Long Papers)}, 2200--2220. Mexico City, Mexico: Association for Computational Linguistics.

\bibitem[{Liu et~al.(2018)Liu, Saleh, Pot, Goodrich, Sepassi, Kaiser, and Shazeer}]{long-doc-efficient}
Liu, P.~J.; Saleh, M.; Pot, E.; Goodrich, B.; Sepassi, R.; Kaiser, L.; and Shazeer, N. 2018.
\newblock Generating Wikipedia by Summarizing Long Sequences.
\newblock In \emph{International Conference on Learning Representations}.

\bibitem[{Liu and Lapata(2019)}]{BertSum-liu-lapata-2019-text}
Liu, Y.; and Lapata, M. 2019.
\newblock Text Summarization with Pretrained Encoders.
\newblock In \emph{Proceedings of the 2019 Conference on Empirical Methods in Natural Language Processing and the 9th International Joint Conference on Natural Language Processing (EMNLP-IJCNLP)}, 3730--3740. Hong Kong, China: Association for Computational Linguistics.

\bibitem[{Liu and Lapata(2021)}]{liu2021noisy-distill}
Liu, Y.; and Lapata, M. 2021.
\newblock Noisy Self-Knowledge Distillation for Text Summarization.
\newblock In \emph{Proceedings of NAACL}, 692--703. Online: Association for Computational Linguistics.

\bibitem[{Liu et~al.(2022)Liu, Ni, Nan, Deb, Zhu, Awadallah, and Radev}]{Pagesum}
Liu, Y.; Ni, A.; Nan, L.; Deb, B.; Zhu, C.; Awadallah, A.~H.; and Radev, D. 2022.
\newblock Leveraging Locality in Abstractive Text Summarization.
\newblock In \emph{Proceedings of the 2022 Conference on Empirical Methods in Natural Language Processing,Abu Dhabi, United Arab Emirates}, 6081--6093. Association for Computational Linguistics.

\bibitem[{Manakul and Gales(2021)}]{manakul-gales-2021-loBART}
Manakul, P.; and Gales, M. 2021.
\newblock Long-Span Summarization via Local Attention and Content Selection.
\newblock In \emph{Proceedings of the 59th Annual Meeting of the Association for Computational Linguistics and the 11th International Joint Conference on Natural Language Processing (Volume 1: Long Papers)}, 6026--6041. Online: Association for Computational Linguistics.

\bibitem[{Mao et~al.(2022)Mao, Wu, Ni, Zhang, Zhang, Yu, Deb, Zhu, Awadallah, and Radev}]{mao-etal-2022-dyle}
Mao, Z.; Wu, C.~H.; Ni, A.; Zhang, Y.; Zhang, R.; Yu, T.; Deb, B.; Zhu, C.; Awadallah, A.; and Radev, D. 2022.
\newblock {DYLE}: Dynamic Latent Extraction for Abstractive Long-Input Summarization.
\newblock In \emph{Proceedings of the 60th Annual Meeting of the Association for Computational Linguistics (Volume 1: Long Papers)}, 1687--1698. Dublin, Ireland: Association for Computational Linguistics.

\bibitem[{Mihalcea and Tarau(2004)}]{mihalcea2004textrank}
Mihalcea, R.; and Tarau, P. 2004.
\newblock {T}ext{R}ank: Bringing Order into Text.
\newblock In \emph{Proceedings of the 2004 Conference on Empirical Methods in Natural Language Processing, Barcelona, Spain}, 404--411. Association for Computational Linguistics.

\bibitem[{Nallapati, Zhai, and Zhou(2017)}]{Nallapati-2017}
Nallapati, R.; Zhai, F.; and Zhou, B. 2017.
\newblock SummaRuNNer: {A} Recurrent Neural Network Based Sequence Model for Extractive Summarization of Documents.
\newblock In \emph{Proceedings of the Thirty-First {AAAI} Conference on Artificial Intelligence, February 4-9, 2017, San Francisco, California, {USA}}, 3075--3081. {AAAI} Press.

\bibitem[{Ou and Lapata(2025)}]{Merella-lapata-2025-context-llm}
Ou, L.; and Lapata, M. 2025.
\newblock Context-Aware Hierarchical Merging for Long Document Summarization.
\newblock In \emph{Findings of the Association for Computational Linguistics: ACL 2025}, 5534--5561. Vienna, Austria: Association for Computational Linguistics.

\bibitem[{Pilault et~al.(2020)Pilault, Li, Subramanian, and Pal}]{extra-super2}
Pilault, J.; Li, R.; Subramanian, S.; and Pal, C. 2020.
\newblock On Extractive and Abstractive Neural Document Summarization with Transformer Language Models.
\newblock In \emph{Proceedings of the 2020 Conference on Empirical Methods in Natural Language Processing, {EMNLP} 2020, Online, November 16-20, 2020}, 9308--9319. Association for Computational Linguistics.

\bibitem[{Radev and McKeown(1998)}]{radev1998generating}
Radev, D.; and McKeown, K. 1998.
\newblock Generating natural language summaries from multiple on-line sources.
\newblock \emph{Computational Linguistics}, 24(3): 469--500.

\bibitem[{Radev et~al.(2004)Radev, Jing, Sty{\'s}, and Tam}]{radev2004centroid}
Radev, D.~R.; Jing, H.; Sty{\'s}, M.; and Tam, D. 2004.
\newblock Centroid-based summarization of multiple documents.
\newblock \emph{Information Processing \& Management}, 40(6): 919--938.

\bibitem[{Reimers and Gurevych(2019)}]{Reimers2019SBERT}
Reimers, N.; and Gurevych, I. 2019.
\newblock Sentence-{BERT}: Sentence Embeddings using {S}iamese {BERT}-Networks.
\newblock In \emph{Proceedings of the 2019 Conference on Empirical Methods in Natural Language Processing and the 9th International Joint Conference on Natural Language Processing (EMNLP-IJCNLP)}, 3982--3992. Hong Kong, China: Association for Computational Linguistics.

\bibitem[{Rohde, Wu, and Liu(2021)}]{HATBART-rohde2021hierarchical}
Rohde, T.; Wu, X.; and Liu, Y. 2021.
\newblock Hierarchical Learning for Generation with Long Source Sequences.
\newblock \emph{CoRR}, abs/2104.07545.

\bibitem[{Rush, Chopra, and Weston(2015)}]{rush-2015-neural-attention}
Rush, A.~M.; Chopra, S.; and Weston, J. 2015.
\newblock A Neural Attention Model for Abstractive Sentence Summarization.
\newblock In \emph{Proceedings of the 2015 Conference on Empirical Methods in Natural Language Processing}, 379--389. Association for Computational Linguistics.

\bibitem[{See, Liu, and Manning(2017)}]{see2017get}
See, A.; Liu, P.~J.; and Manning, C.~D. 2017.
\newblock Get to the Point: Summarization with Pointer-Generator Networks.
\newblock In \emph{Proceedings of the 55th Annual Meeting of the Association for Computational Linguistics}, 1073--1083.

\bibitem[{Shen et~al.(2024)Shen, Lam, Ma, and Wang}]{text-align-UNBALANCED-OPTIMAL-TRANSPORT}
Shen, X.; Lam, W.; Ma, S.; and Wang, H. 2024.
\newblock Joint learning of text alignment and abstractive summarization for long documents via unbalanced optimal transport.
\newblock \emph{Natural Language Engineering}, 30(3): 525–553.

\bibitem[{Sutskever, Vinyals, and Le(2014)}]{Seq-to-seq}
Sutskever, I.; Vinyals, O.; and Le, Q.~V. 2014.
\newblock Sequence to Sequence Learning with Neural Networks.
\newblock In \emph{Advances in Neural Information Processing Systems 27: Annual Conference on Neural Information Processing Systems 2014, December 8-13 2014, Montreal, Quebec, Canada}, 3104--3112.

\bibitem[{Tay et~al.(2020)Tay, Bahri, Yang, Metzler, and Juan}]{sink-horn-attention}
Tay, Y.; Bahri, D.; Yang, L.; Metzler, D.; and Juan, D. 2020.
\newblock Sparse Sinkhorn Attention.
\newblock In \emph{Proceedings of the 37th International Conference on Machine Learning, {ICML} 2020, 13-18 July 2020, Virtual Event}, volume 119 of \emph{Proceedings of Machine Learning Research}, 9438--9447. {PMLR}.

\bibitem[{Tay et~al.(2022)Tay, Dehghani, Bahri, and Metzler}]{tay2022efficient}
Tay, Y.; Dehghani, M.; Bahri, D.; and Metzler, D. 2022.
\newblock Efficient Transformers: A Survey.
\newblock \emph{ACM Comput. Surv.}, 55(6).

\bibitem[{Vaswani et~al.(2017)Vaswani, Shazeer, Parmar, Uszkoreit, Jones, Gomez, Kaiser, and Polosukhin}]{vaswani2017attention}
Vaswani, A.; Shazeer, N.; Parmar, N.; Uszkoreit, J.; Jones, L.; Gomez, A.~N.; Kaiser, L.~u.; and Polosukhin, I. 2017.
\newblock Attention is All you Need.
\newblock In Guyon, I.; Luxburg, U.~V.; Bengio, S.; Wallach, H.; Fergus, R.; Vishwanathan, S.; and Garnett, R., eds., \emph{Advances in Neural Information Processing Systems}, volume~30, 5998--6008. Curran Associates, Inc.

\bibitem[{Witbrock and Mittal(1999)}]{WitbrockM99}
Witbrock, M.~J.; and Mittal, V.~O. 1999.
\newblock Ultra-Summarization: {A} Statistical Approach to Generating Highly Condensed Non-Extractive Summaries.
\newblock In \emph{Proceedings of the 22nd Annual International {ACM} {SIGIR} Conference on Research and Development in Information Retrieval, Berkeley, CA, USA}, 315--316. ACM.

\bibitem[{Xiao et~al.(2022)Xiao, Beltagy, Carenini, and Cohan}]{xiao-etal-2022-primera}
Xiao, W.; Beltagy, I.; Carenini, G.; and Cohan, A. 2022.
\newblock {PRIMERA}: Pyramid-based Masked Sentence Pre-training for Multi-document Summarization.
\newblock In \emph{Proceedings of the 60th Annual Meeting of the Association for Computational Linguistics (Volume 1: Long Papers)}, 5245--5263. Dublin, Ireland: Association for Computational Linguistics.

\bibitem[{Xiao and Carenini(2019)}]{Extract-super1}
Xiao, W.; and Carenini, G. 2019.
\newblock Extractive Summarization of Long Documents by Combining Global and Local Context.
\newblock In \emph{Proceedings of the 2019 Conference on Empirical Methods in Natural Language Processing and the 9th International Joint Conference on Natural Language Processing, {EMNLP-IJCNLP} 2019, Hong Kong, China, November 3-7, 2019}, 3009--3019. Association for Computational Linguistics.

\bibitem[{Zaheer et~al.(2020)Zaheer, Guruganesh, Dubey, Ainslie, Alberti, Ontanon, Pham, Ravula, Wang, Yang, and Ahmed}]{Zaheer}
Zaheer, M.; Guruganesh, G.; Dubey, K.~A.; Ainslie, J.; Alberti, C.; Ontanon, S.; Pham, P.; Ravula, A.; Wang, Q.; Yang, L.; and Ahmed, A. 2020.
\newblock Big Bird: Transformers for Longer Sequences.
\newblock In \emph{Advances in Neural Information Processing Systems}, volume~33, 17283--17297. Curran Associates, Inc.

\bibitem[{Zhang et~al.(2021)Zhang, Gong, Shen, Li, Lv, Duan, and Chen}]{poolingformer}
Zhang, H.; Gong, Y.; Shen, Y.; Li, W.; Lv, J.; Duan, N.; and Chen, W. 2021.
\newblock Poolingformer: Long Document Modeling with Pooling Attention.
\newblock In \emph{Proceedings of the 38th International Conference on Machine Learning, {ICML} 2021, Virtual Event}, volume 139 of \emph{Proceedings of Machine Learning Research}, 12437--12446. {PMLR}.

\bibitem[{Zhang et~al.(2020)Zhang, Zhao, Saleh, and Liu}]{pegasus-zhang-2020}
Zhang, J.; Zhao, Y.; Saleh, M.; and Liu, P.~J. 2020.
\newblock {PEGASUS:} Pre-training with Extracted Gap-sentences for Abstractive Summarization.
\newblock In \emph{Proceedings of the 37th International Conference on Machine Learning, {ICML} 2020, 13-18 July 2020, Virtual Event}, volume 119 of \emph{Proceedings of Machine Learning Research}, 11328--11339. {PMLR}.

\bibitem[{Zheng and Lapata(2019)}]{ZhengL19}
Zheng, H.; and Lapata, M. 2019.
\newblock Sentence Centrality Revisited for Unsupervised Summarization.
\newblock In \emph{Proceedings of the 57th Conference of the Association for Computational Linguistics, Florence, Italy, 2019, Volume 1: Long Papers}, 6236--6247. Association for Computational Linguistics.

\end{thebibliography}

\end{document}